\documentclass[prl,twocolumn,superscriptaddress,nofootinbib,aps,10pt,preprintnumbers,
longbibliography]{revtex4-1}

\usepackage[colorlinks=true,linkcolor=black,citecolor=blue,urlcolor=blue, pdfborder={0 0 0}]{hyperref}
\usepackage{mathtools}
\usepackage[utf8]{inputenc}
\usepackage{color}
\usepackage[table,xcdraw,dvipsnames]{xcolor}
\usepackage{multirow}
\usepackage[normalem]{ulem}    

\def\eqs#1#2{{Eqs.~(\ref{#1}) and (\ref{#2})}}

\def\fig#1{{Fig.~\ref{#1}}}

\def\Table#1{{Table~\ref{#1}}}

\newcommand{\Sec}[1]{ \medskip \noindent {\sl \bfseries #1}}

\definecolor{ultramarine}{RGB}{0,32,96}

\renewcommand{\bar}{\overline}

\newcommand{\gae}{g_{ae}}
\newcommand{\gag}{g_{a\gamma}}

\newcommand{\beq}{\begin{equation}}
\newcommand{\eeq}{\end{equation}}
\newcommand{\bea}{\begin{eqnarray}}
\newcommand{\eea}{\end{eqnarray}}

\newcommand{\eqn}[1]{Eq.~(\ref{#1})}

\def \lsim{\mathrel{\vcenter
     {\hbox{$<$}\nointerlineskip\hbox{$\sim$}}}}
\def \gsim{\mathrel{\vcenter
     {\hbox{$>$}\nointerlineskip\hbox{$\sim$}}}}
     
\begin{document}

\preprint{DESY 20-106}
\title{Solar axions cannot explain the 
XENON1T excess
}

\newcommand{\affBARRY}{{\small \it Physical Sciences, Barry University, 11300 NE 2nd Ave., Miami Shores, FL 33161, USA}}
\newcommand{\affINFN}{{\small \it INFN, Laboratori Nazionali di Frascati, C.P.~13, 100044 Frascati, Italy}} 
\newcommand{\affDESY}{{\small \it Deutsches Elektronen-Synchrotron DESY, Notkestra{\ss}e 85, 
D-22607 Hamburg,  Germany}}
\newcommand{\affUB}{{\small \it Dept.~de F\'{\i}sica Qu\`antica i Astrof\'{\i}sica, Institut de Ci\`encies del Cosmos (ICCUB), Universitat de Barcelona, Mart\'i i Franqu\`es 1, E-08028 Barcelona, Spain}}

\author{Luca Di Luzio}
\email{luca.diluzio@desy.de}
\affiliation{\affDESY}

\author{Marco Fedele}
\email{marco.fedele@icc.ub.edu}
\affiliation{\affUB}

\author{Maurizio Giannotti}
\email{MGiannotti@barry.edu}
\affiliation{\affBARRY}

\author{Federico Mescia}
\email{mescia@ub.edu}
\affiliation{\affUB}

\author{Enrico Nardi}
\email{enrico.nardi@lnf.infn.it}
\affiliation{\affINFN}

\begin{abstract}
We argue that the interpretation in terms 
of solar axions
of the 
recent XENON1T excess is not 
tenable when confronted with astrophysical observations of stellar evolution. 
We discuss the reasons why the emission of a flux of solar axions 
sufficiently intense to explain the anomalous data would 
radically alter the distribution of certain type of stars in 
the color-magnitude diagram 
in first place,  and would also clash with a 
certain number of other astrophysical observables.
Quantitatively,
the significance of the discrepancy ranges from $3.3\sigma$ 
for the rate of period change 
of pulsating white dwarfs, 
and exceeds $19\sigma$ for the $R$-parameter and for $M_{I,{\rm TRGB}}$.
\end{abstract}

\maketitle

\Sec{Introduction.}
The XENON1T collaboration \cite{Aprile:2020tmw} has  reported an excess in low-energy electronic recoil data below 7~keV and peaking around 
2-3~keV.
The collaboration cautions that the excess could  be 
due to  an unaccounted background from 
$\beta$ decays due to a trace amount of tritium,
but they also explore the possibility 
that the signal is due to different types of new physics. 
The most intriguing interpretation, which also provides the best fit to the data, is given in terms of solar axions, favoured over the background-only hypothesis at the $3.5\sigma$ level. \hfill\break\indent
Three production mechanisms  contribute to the solar axion flux: $i)$ 
Atomic recombination and deexcitation, Bremsstrahlung, and Compton (ABC) interactions \cite{Redondo:2013wwa} 
that are controlled by  the axion-electron coupling $\gae$, 
$ii)$  Primakoff conversion of photons into axions
\cite{Pirmakoff:1951pj} 
induced by the axion-photon coupling $\gag$, 
$iii)$  axion emission in the M1 nuclear transition 
of $^{57}$Fe~\cite{Moriyama:1995bz}
that produces  mono-energetic 14.4 keV axions, and is 
controlled by and effective  
axion-nucleon coupling $g_{an}^{\rm eff}$.
Since this last process cannot   play  any role in producing events
below 10~keV, we will not include in our analysis astrophysical 
observables sensitive to $g_{an}^{\rm eff}$.
Conversely, axions produced through   $i)$ and $ii)$  
feature a wide spectrum peaking around a few keV. 
The production rates are independent of the axion  mass
for $m_a \lesssim 100\,$eV. 
As regards  detection, electron recoils occur via the axio-electric 
effect which is controlled by $\gae$.
Because of this, and because the location of the
peak around 2-3 keV corresponds roughly to the maximum of 
the axion energy spectrum
for the ABC processes, 
the Primakoff and $^{57}$Fe components are both allowed to be absent 
as long as there is a nonzero ABC component. 
This selects $\gae$ as the
crucial coupling to attempt to explain the data in terms of the 
QCD axion~\cite{Peccei:1977hh,Peccei:1977ur,
Weinberg:1977ma,Wilczek:1977pj}.\footnote{Our results 
apply also to explanations based on 
generic axion-like particles, for which there is no 
theoretical relation between $m_a$ and the  coupling strengths, and that
are unrelated to the strong CP problem.}
Taken at face value the strength of the 
XENON1T excess requires $\gae\gsim 10^{-12}$, corresponding  
to an axion decay constant  $f_a\lsim 10^8\,$GeV, and in turn to  
an axion mass $m_a\gsim 0.06\,$eV. 
However, astrophysical considerations  indicate that such a large value of $\gae$ is not tenable, as stellar evolution would be 
drastically affected by the 
exceedingly large energy losses
via axion emission.
The strategy that we will follow consists in assuming that  
$\gae$ and $\gag$ lie in the 90\% C.L. 
regions resulting from the XENON1T fit~\cite{Aprile:2020tmw}. 
We will then estimate the effects of 
extra energy losses on a  set of astrophysical observables 
related to Red Giants Branch (RGB) and Horizontal Branch (HB) stars, 
and to White Dwarfs (WDs).

\Sec{Astrophysical~observables~and~axion~couplings.}
The axion interactions with photons and electrons read
\beq 
\label{eq:Laxion}
\mathcal{L}_{\rm int} = 
\frac{1}{4} g_{a\gamma} a F_{\mu\nu} \tilde F^{\mu\nu} 
+ g_{ae} \frac{\partial_\mu a}{2m_e} 
\bar e \gamma^\mu \gamma_5 e\, , 
\eeq
where the couplings 
can be related to model-dependent dimensionless coefficients as 
$g_{a\gamma} = \frac{\alpha}{2\pi} \frac{C_{a\gamma}}{f_a}$ 
and 
$g_{ae} = C_{ae} \frac{m_e}{f_a}$.   
In benchmark axion models 
$C_{a\gamma}$ and $C_{ae}$ are  typically of $\mathcal{O}(1)$, 
although  strong enhancements/suppressions are possible
in specific cases~\cite{DiLuzio:2016sbl,DiLuzio:2017pfr,DiLuzio:2017ogq,Bjorkeroth:2019jtx,DiLuzio:2020wdo}. 
In the following, 
we will adopt the notation
$g_{\gamma 10} \equiv g_{a\gamma} \times \left(10^{10} \, \text{GeV}\right)$ 
and 
$g_{e 13} \equiv g_{ae} \times 10^{13}$.
Axions with couplings as large as  
$g_{e 13} \sim 10 $, $g_{\gamma 10} \sim 1 $ would be 
abundantly produced in several types of stars without being trapped, and thus would efficiently 
drain energy from the star cores

Astrophysical considerations have been systematically used to place severe bounds on light,  
weakly interacting particles, such as neutrinos and axions~\cite{Raffelt:1996wa}. 
Noticeably, a set of anomalous 
observations have recently led to speculations that new physics is 
at play \cite{Hoof:2018ieb,DiVecchia:2019ejf,DiLuzio:2020wdo}, and 
the axion case
appears especially compelling~\cite{Giannotti:2017hny,Giannotti:2015kwo}. 
The most  effective observables to constrain $\gae$ and $\gag$ 
are described below.

{$\bullet\ $\it Tip of RGB stars in globular cluster.}
We denote by $M_{I,{\rm TRGB}}$ 
the luminosity of the tip of the RGB  in globular clusters. 
RG stars are characterized by a He core and a burning H shell. 
During the RGB evolution, the ashes from the burning shell increase 
the He core mass, while the star luminosity (determined by
equilibrium at the surface of the 
He core between  thermal pressure supporting the non-degenerate envelope
against the gravity pull from the core) keeps growing.
The process continues until the core reaches sufficiently large 
temperatures and  densities 
($T \sim 10^8\,$K, $\rho=10^6\,$g\,cm$^{-3}$)
to ignite He, an event known as the He-flash. 
At this stage the star has reached its maximum luminosity $M_{I,{\rm TRGB}}$,  
after which it shrinks and moves to the HB. 
If an additional core-cooling mechanism were at play,  He ignition would be delayed, 
the core would accrete a larger mass, and the star would reach 
higher luminosities. Therefore, measurements of $M_{I,{\rm TRGB}}$ 
allow to test the rate of cooling during the RGB phase.  
The method is particularly effective for constraining $\gae$ 
since in RG cores axions can be efficiently produced 
via electron bremsstrahlung.
The most recent analyses~\cite{Viaux:2013lha,Straniero:2018fbv,Diaz:2019kim} 
have derived comparable constraints.
Here we adopt the result of the analysis of the Large Magellanic Cloud 
in Ref.~\cite{Freedman:2019jwv,Freedman:2020dne} which provides the most conservative bound
$M_{I,{\rm TRGB}} = -4.047 \pm 0.045\ {\rm mag}$. 
In terms of $\gae$  this observable can be written as~\cite{Viaux:2013lha,Capozzi:2020cbu}
\begin{align}
\label{eq:MITRGB_theo}
M_{I,{\rm TRGB}}^{\rm theo} & = - 4.08 
\nonumber 
\\
- 0.25 & \bigg(\sqrt{g_{e13}^2 + 0.93} 
- 0.96 - 0.17 g_{e13}^{1.5}\bigg) \,, 
\end{align}
which results from an analytic fit to 
ten evolutionary track points reaching close 
to the RGB tip obtained from numerical simulations~\cite{Viaux:2013hca}, 
and corresponding to values of $g_{e13}$
up to 9~\cite{Viaux:2013lha}.
The associated theoretical uncertainty
is $\sigma^2 =
{0.039^2 + (0.046 + 0.012 g_{e13})^2}$~\cite{Viaux:2013lha}.

{$\bullet\ $\it $R$-parameter.}
After He ignition the RG core expands and its density decreases 
by  two orders of magnitude.
The star migrates to the HB  
and remains supported by He burning in a non-degenerate core.
The ratio $R=N_{\rm HB}/N_{\rm RGB}$ between the number of stars in 
in globular clusters in the HB and in the upper portion of the RGB 
directly measures the duration of  He burning  in the HB phase.
The value $R = 1.39 \pm 0.03$ was obtained in Ref.~\cite{Ayala:2014pea}
from the analysis of 39 clusters.
The duration of  the HB phase can be affected by $\gae$-related processes 
both directly and indirectly. 
If $\gae$ is sufficiently large, axion emission would directly produce 
extra cooling of the He core. The 
star self-regulates by slightly contracting,  the 
core temperature increases speeding up the He burning rate. 
Once the He fuel is exhausted, the star turns into a WD.
The indirect effect is related to the 
growth of the degenerate He core during the RGB phase 
previously discussed. 
HB stars would  unavoidably inherit a more massive core from the parent RGs, 
resulting in an increased He burning rate
to contrast the larger gravitational pull, and
shortening further the duration of the HB phase. 
Note that the indirect effect of $\gae$ 
is so important  that for 
$g_{e13} \sim 15$ it would  suffice to depopulate 
almost completely the HB in the 
Color-Magnitude Diagram (CMD)
$(R\approx 0)$.
Cooling of HB stars can also proceed via the  Primakoff 
effect $\propto \gag^2$, which is particularly efficient at the 
typical temperatures and densities of HB cores ($T \sim 10^8\,$K, $\rho=10^4\,$g\,cm$^{-3}$).
For sufficiently large values of $\gag$,  $R$  can still decrease 
well below the observed values even when $\gae \approx 0$. 
Hence an accurate determination of this observable allows 
to probe the axion coupling to both photons and electrons.
In terms of $\gae$ and $\gag$ 
the $R$-parameter 
can be written as~\cite{Giannotti:2015kwo,Straniero:2015nvc}
\bea
\label{eq:Rparam_theo}
\!\!\!\!\!\!\!\! R^{\rm theo} &=& 7.33Y - 0.095\sqrt{21.86 + 21.08 g_{\gamma 10}} \nonumber\\
&+& 0.02 -1.61\delta \mathcal{M}_c - 0.005 g_{e13}^2 \,, \\
\nonumber
\!\!\!\!\!\!\!\! \delta \mathcal{M}_c &=& 0.024 \bigg(\sqrt{g_{e13}^2 + 1.23^2} 
- 1.23 - 0.138\,g_{e13}^{1.5} \bigg),
\eea
where $\delta \mathcal{M}_c$
is the change in the He-core mass, and $Y \simeq 0.255 \pm 0.002$ is the primordial He abundance.
The relative errors on $\delta \mathcal{M}_c$, which  
represents the main  theoretical uncertainty from astrophysics.
and the one on $Y$, are of the same order.
Hence, due to the larger coefficient multiplying $Y$, 
the  uncertainty from  $\delta \mathcal{M}_c$   can be neglected. 
Similarly to \eqn{eq:MITRGB_theo} this  expression is derived from an analytic 
fit to evolutionary track points  calculated with stellar evolutionary codes modified to account for axion emission~\cite{Giannotti:2015kwo,Straniero:2015nvc},
and thus it is  quantitatively reliable
up to  values of $\gae$
not much larger than those corresponding to the last point fitted  
(for definiteness  $g_{e13} \sim 9$).
Thus, we will not input into these expressions 
the much larger XENON1T  values $g_{e13}\sim 30$. Rather,  
very conservatively,  we will limit ourselves to estimate the 
tension between the observed values of  $M_{I,{\rm TRGB}}$ and $R$,  and the values resulting from  Eqs.~\eqref{eq:MITRGB_theo} and \eqref{eq:Rparam_theo} when evaluated  at $g_{e13} \sim 9$ ($\gag\approx 0$). 
As regards values of $\gag$ too large to be used 
in \eqn{eq:Rparam_theo},  they can be directly constrained   
from the lifetime of  HB stars which, 
in the presence of extra cooling,
scales as $\sim L_0/(L_0+L_a)$ with $L_0$ ($L_a$) the  
standard (axion) core luminosity~\cite{Raffelt:1996wa}.
Hence for  $g_{\gamma10}\gsim 1 $, rather than~\eqn{eq:Rparam_theo}, 
we will 
use 
\beq
\label{eq:Rparam_theo2}
R^{\rm theo} \approx \dfrac{a^2}{a+b\, g_{\gamma 10}^2} \, ,
\eeq
with $a=(6.26 Y - 0.12)$ and $b=0.41$~\cite{Ayala:2014pea}.
Note that \eqn{eq:Rparam_theo2} 
neglects both direct and indirect 
effects of $\gae$ on HB and RGB stars, and 
hence it would also yield conservative limits.

{$\bullet\ $\it White Dwarf luminosity function.}
The third observable we consider is the distribution of WDs as a function of their luminosity (WDLF).
The WDLF measures  the WD cooling efficiency, and thus allows to place strong bounds on new exotic 
cooling processes, including axion emission (see Ref.~\cite{Isern:2020non} for a review).
WDs are compact objects whose hydrostatic equilibrium is supported by electron degeneracy pressure, 
hence  axion emission from WDs would dominantly depend on $\gae$. 
Here we will use the bound $g_{e13}^{\rm WDLF} \leq 2.8$
obtained in Ref.~\cite{Bertolami:2014wua}.

{$\bullet\ $\it Rate of period change of  WD variables.}
WD variables (WDV)  
are WDs whose luminosity varies periodically, 
with a period $\Pi$ ranging from a few to several minutes.
Because the oscillation period depends on the luminosity, a 
secular change   of the period $\dot{\Pi}$
tracks the rate of decrease of the star luminosity.
To a very good approximation
$\dot \Pi/\Pi$ is  proportional to the cooling rate  $\dot T/T$, hence 
a  measurement of  $\dot{\Pi}$
allows to constrain possible sources of extra cooling  
(see Ref.~\cite{Corsico:2019nmr} for a review).
Here we consider four  WDVs:
G117-B15A~\cite{Corsico:2012ki}, R548~\cite{Corsico:2012sh}, L19-2~\cite{Corsico:2016okh} 
(for two pulsation modes),
and PG1351+489~\cite{Battich:2016htm}.  
We list in  \Table{table:Obs}
the corresponding measured values of $\dot \Pi/\Pi$. 
Theoretically, the  rate of change in the WD pulsating period as a function of $g_{e13}$ 
can be  parametrized as~\cite{Giannotti:2017hny}:
$\dot{\Pi}_{\rm WD_i}^{\rm theo} = a_i + b_i\ g_{e13}^2$,
where $a_i$ and $b_i$ are parameters specific for each WD.

\begin{figure}[t]
\centering
\includegraphics[width=8cm]{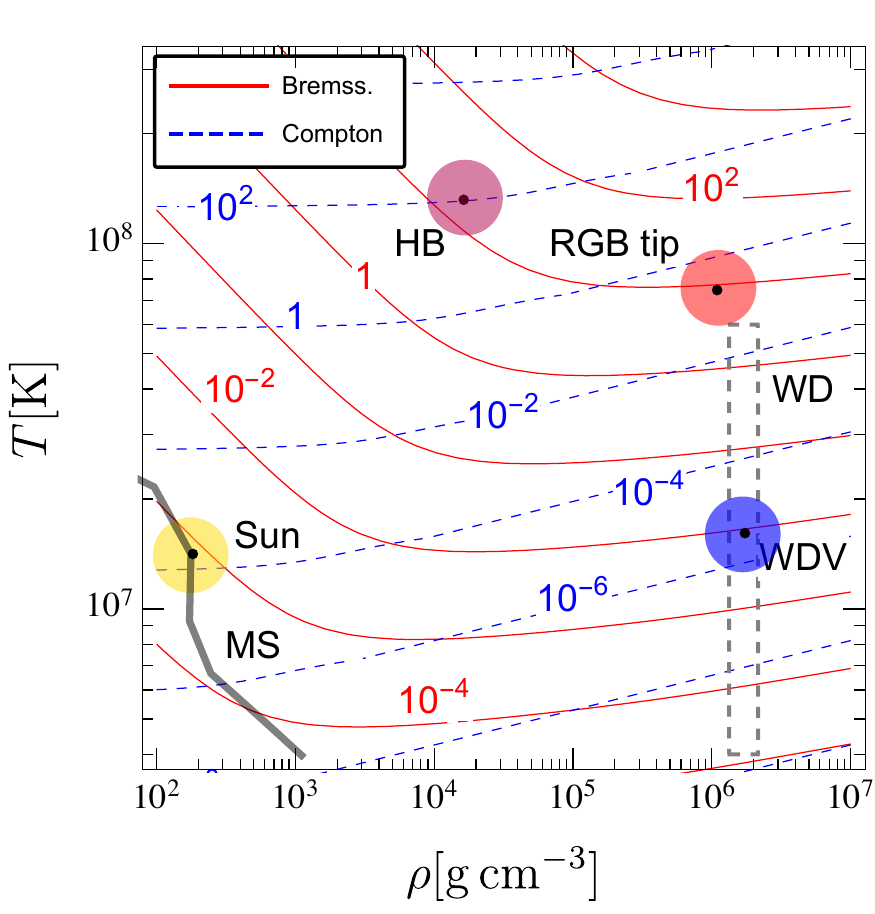}
\caption{Contours of the axion energy-loss rates per unit mass,
in erg$\,$g$^{-1}$s$^{-1}$, for a pure He plasma
and $g_{e13}=4.3$. 
\label{fig:bremscompt}}
\end{figure}

\Sec{XENON1T vs.~Astrophysics.}
\fig{fig:bremscompt} shows contours of the axion energy-loss 
rates per unit mass in a temperature vs.~density plane, for a pure He plasma. 
Contour iso-lines for energy-loss due to  Compton (dashed blue)  
and Bremsstrahlung (solid red) processes,  which 
are controlled by $g_{ae}$, are also shown.  
For reference, we have fixed 
$g_{e13}=4.3$, which corresponds 
to the RGB bound from M5~\cite{Viaux:2013lha}.    
Energy loss rates for different values of $g_{ae}$ can be easily obtained recalling
that they scale as $g_{ae}^2$.
The labelled disks in the figure show the position of the RGB tip 
and of a typical HB star
(of mass $0.8 M_\odot$) 
  and a range of WDs with luminosities varying 
  from $5\times 10^{-4} $ to $ 5\times 10^{-1}\,L_{\odot}$  (dashed gray rectangle).
The blue disk indicates the temperature/density  of a typical WD variable~\citep{Corsico:2019nmr}.
The location of the Sun 
is marked with a yellow disk on top of the 
broken gray line 
which locates  Main Sequence (MS) stars of different masses.
Note that since MS stars, including the Sun, 
are supported by H burning cores, their position with 
respect to the energy loss iso-lines for the He plasma 
is approximate, and slightly shifted towards larger rates.
The picture shows clearly that the Sun is a relatively 
faint axion emitter with respect to other stellar objects, 
so that  values $g_{e13} \gsim 10$ as required 
to account for the XENON1T excess
would unavoidably turn other stars into bright `axion lighthouses'. 
The RGB would extend to  higher luminosities than the  
ones observed, and the decreased duration of the He burning phase  
would depopulate the HB, to the point that for smaller clusters with  
relatively few stars, already for 
$g_{e13} \sim 15$ we would expect  $R\approx 0$.  
In short, regardless of other details, a value  $g_{e13}\sim 30$
would definitely destroy the agreement between stellar evolution 
models and the observed CMD. 

\begin{figure}[t]
\centering
\includegraphics[width=8cm]{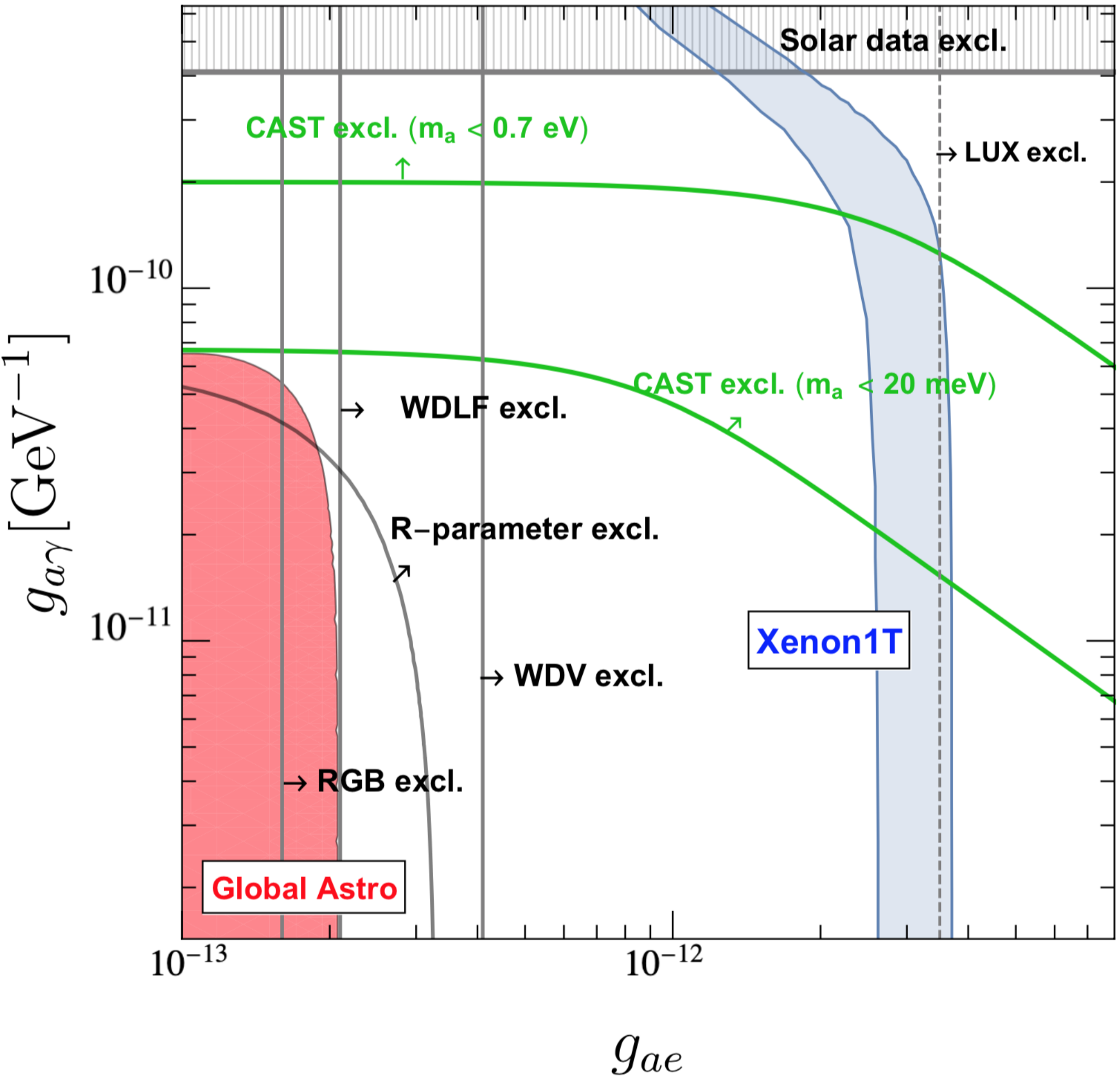}
\caption{XENON1T $90\%$ C.L.~fit (blue region). 
$3\sigma$ exclusion limit from solar data (grey hatched region).
$2 \sigma$ LUX  limit (grey dashed line) and CAST limits 
for $m_a < 20\,$meV and  $m_a < 0.7\,$eV (green lines). 
Individual $2\sigma$ limits from 
$R$-parameter, TRGB, WDLF,  WDVs (grey lines) and $2\sigma$
global bound from astrophysics (red region). 
\label{fig:boundsgae}
}
\end{figure}

\Sec{Quantifying the tension.}
The projections of the XENON1T $90\%$~C.L. best fit region 
onto the $(g_{ae},\,g_{a\gamma})$, $(g_{ae},\, g_{an}^{\rm eff})$
and $(g_{ae}g_{a\gamma},\, g_{ae}g_{an}^{\rm eff})$ planes 
are given in Fig.~8 of Ref.~\cite{Aprile:2020tmw}.
Since   only  $g_{ae}$ and $g_{a\gamma}$ can be responsible 
for the anomalous XENON1T data below 7\,keV, we focus on the 
best fit region for these two couplings, that corresponds 
the blue band  in \fig{fig:boundsgae}. 
In the figure we also show the $2\sigma$ limits on $g_{ae},\,g_{a\gamma}$ 
obtained from each single astrophysical observable, as well as the result of 
a global fit to the entire set of stellar cooling data. 
The curve depicting the CAST~\cite{CAST2009} limit 
in the $(g_{ae},\,g_{a\gamma})$ plane 
in Ref.~\cite{Aprile:2020tmw} was taken from
Ref.~\cite{Barth:2013sma}. We  update this bound with  
the most recent CAST results~\cite{Anastassopoulos:2017ftl} 
which, in the $\gae\simeq0$ limit, and for 
$m_a\lesssim 20\,$meV ($ m_a\lesssim 0.7\,$eV),
correspond to $\gag < 0.66\, (2.0) \times 10^{-10}\,$GeV$^{-1}$.
These  limits are represented in \fig{fig:boundsgae} by 
the two green lines, in which we have folded in the 
effects of a non-zero $\gae$ that   would increase 
the production of solar axions and
strengthen the bounds. 
The vertical dashed line is 
 LUX limit~\cite{Akerib:2017uem}. 
The grey horizontal  line  
at $g_{\gamma10}=4.1$ corresponds to the limit 
from a global fit to solar data, which includes the 
measured flux of $^8$B and $^7$Be neutrinos as well as  
additional data inferred from helioseismology 
observations~\cite{Vinyoles:2015aba}.  
This is about a factor of two stronger than 
the bound labeled  ``solar $\nu$" in the  upper panel 
of Fig.~8  in Ref.~\cite{Aprile:2020tmw} which is   
taken from  Ref.~\cite{Gondolo:2008dd}.\footnote{For values 
of the couplings allowed by astrophysics the solar axion 
luminosity $L_{a}$ is a negligible fraction of the total luminosity,
 for example $L_a \approx 1.85 
\times 10^{-3} g^2_{\gamma10} L_\odot$ for Primakoff emission~\cite{Gondolo:2008dd}.
Hence, effects on the Sun lifetime are also negligible.} 
To assess quantitatively the discrepancy
between the values of $\gae$ and $\gag$ 
needed to reproduce the XENON1T excess we proceed as follows: 
we first extract the allowed ranges
from the $90\%$ C.L.~region of Ref.~\cite{Aprile:2020tmw}
not excluded by solar data
(the blue area in \fig{fig:boundsgae}).
This region can be parametrized by means of 
an effective coupling~\cite{DiLuzio:2020wdo}
\beq
\label{eq:XenonRel}
\bar{g}_{e13}^4 = g_{e13}^2 (g_{e13}^2 + 200 g_{\gamma10}^2)\,. 
\eeq
The $90\%$ C.L.~($68\%$ C.L.) region of XENON1T 
is then well represented by  the range
{$\bar{g}_{e13} \in [26, 37]$}
{($\bar{g}_{e13} \in[28, 35]$)}.
Varying  $g_{ae}$ and $g_{a\gamma}$ 
with the constraint that {$\bar{g}_{e13}$}
remains within this range, we estimate the range of  values 
for the astrophysical observables implied by the XENON1T data,  
and we confront them with the measured values.
\begin{table}[t]
\renewcommand{\arraystretch}{1.5}
\begin{center}
\footnotesize
\begin{tabular}{|c|c|c|c|}
\hline
Observable &
 Measured
&
Expected
& Tension \\
\hline
\hline
$R$-parameter & $1.39 \pm 0.03$ 
& $\leq 0.83\; (g_{e13}=9)$ & $19\sigma^{\star}$  \\
\hline
$M_{I,{\rm TRGB}}^{\rm LMC}\ [\text{mag}]$ & $-4.047 \pm 0.045$
& $\leq - 4.92\; (g_{e13}=9)$ & $19\sigma^{\star}$ \\
\hline
$g_{e13}^{\rm WDLF}$ & $\leq 2.8 \ (3 \sigma)$ 
& $29.7 \pm 4.8$ & $5.6\sigma$ \\
\hline
$\dot{\Pi}_{\rm L19-2}^{(113)}$ & $3.0 \pm 0.6$ 
& $57 \pm 16$ & $3.4\sigma$  \\
\hline
$\dot{\Pi}_{\rm L19-2}^{(192)}$ & $3.0 \pm 0.6$ 
& $95 \pm 27$ & $3.4\sigma$ \\
\hline
$\dot{\Pi}_{\rm PG1351+489}$ & $200 \pm 90$ 
& $19620 \pm 5730$ & $3.4\sigma$ \\
\hline
$\dot{\Pi}_{\rm G117-B15A}$ & $4.2 \pm 0.7$ 
& $113 \pm 33$ & $3.3\sigma$ \\
\hline
$\dot{\Pi}_{\rm R548}$ & $3.3 \pm 1.1$ 
& $87 \pm 25$ & $3.3\sigma$ \\
\hline
\end{tabular}
\end{center}
\caption{
Measured values of astrophysic observables 
and expected ranges, for $\gae,\,\gag$ falling   
within the $1\sigma$ region of the XENON1T fit
{($\bar{g}_{e13} \in [28, 35]$)}. 
$\dot{\Pi}_{\rm WD_i}$ are in  units of $[10^{-15} s/s]$. 
For $R$ and $M_{I,{\rm TRGB}}$ the expected regions and tensions 
correspond to 
{$g_{e13} =\bar{g}_{e13}(\gag=0) \geq 9$}
(see text). 
}
\label{table:Obs}
\end{table}
Our results are collected in \Table{table:Obs}.
For each observable, the tension given in the fourth column  
is evaluated by dividing the difference between 
the value implied by the  XENON1T data and the astrophysical determination,  by  the total uncertainty.
Given that the statistical distributions are at best only approximately known,
these results are only indicative, and have no rigorous Gaussian meaning.
It is apparent  that the large $\gae$ 
required to fit the XENON1T excess
are in strong conflict with all the astrophysical observables. 
The discrepancy is at the level  of $\sim 3.4\sigma$
for the WDVs cooling rates (last five rows in the Table), 
and reaches $\sim 6\sigma$ for the WDLF in the third row. 
As regards the first two rows,  
the  expected values of $R^{\rm theo}$ and of $M_{I,{\rm TRGB}}^{\rm theo}$
reported in the table are obtained respectively from \eqn{eq:Rparam_theo}  
and \eqn{eq:MITRGB_theo} by setting $g_{e13} = 9$, rather than 
by inserting the much larger values $g_{e13}\sim 30$ 
needed to account for the XENON1T data. 
This is a precautionary procedure that we have adopted 
to avoid estrapolating~\eqs{eq:MITRGB_theo}{eq:Rparam_theo} to 
values of $\gae$ for which the quantitative accuracy of these 
parametrizations cannot be easily assessed.
We have then marked with a $\star$ the corresponding 
tensions. We expect that values of the observables in agreement with 
the XENON1T solar axion fit would result in much larger tensions. 
For example,   already for $g_{e13} \approx 15$~\eqn{eq:Rparam_theo} would yield 
$R\approx 0$, corresponding to a complete depopulation of the HB,  
and $46\sigma$ away from observations.

\Sec{Conclusions.}
In this work, we have explained why astrophysical observations 
firmly exclude that solar axions could account for the XENON1T excess.
Other explanations based on solar production
of new light particles 
or on modifications of neutrino properties  
(such as a neutrino magnetic moment) are also prone to severe 
astrophysical constraints, and as long as the  corresponding 
new physics processes would also occur in RG, HB and WD stellar cores,
they can likewise be excluded.\footnote{Astrophysical 
constraints could only be evaded in exotic models in which the 
couplings strongly depend on the stellar environment, like the core density and temperature, see e.g.~\cite{Redondo:2008tq}.}

If it will be eventually found 
that the tritium background 
or other systematic effects~\cite{Dessert:2020vxy,Szydagis:2020isq}
are not responsible for the excess,
other mechanisms involving either absorption or 
scattering of new particles of non-solar origin 
off target electrons~\cite{Takahashi:2020bpq,Kannike:2020agf,
Alonso-Alvarez:2020cdv,Boehm:2020ltd,Fornal:2020npv}, 
although less compelling than the QCD axion, 
might still provide viable explanations for the XENON1T data.

\Sec{Note added.}
After completing this letter, Refs.~\cite{Gao:2020wer,Dent:2020jhf} 
appeared claiming that besides the axio-electric effect, also the inverse 
Primakoff process can contribute to the detection 
of solar axions by XENON1T. 
This would relax the 
best fit region towards lower values of $\gae$ at the cost of 
increasing $\gag$.
This can relax the tension with astrophysical bounds, however,   
using the results of Ref.~\cite{Gao:2020wer,Dent:2020jhf} 
we have verified that the discrepancy with the $R$-parameter 
remains at least at the level of~$8\sigma$.

\Sec{Acknowledgments.}
We thank Axel Lindner for useful comments. LDL is supported by the Marie Sk\l{}odowska-Curie Individual Fellowship grant AXIONRUSH (GA 840791).
MF and FM are supported by MINECO grant FPA2016-76005-C2-1-P, Maria de Maetzu program grant MDM-2014-0367 of ICCUB and 2017 SGR 929. EN is supported by the INFN Iniziativa Specifica, Theoretical Astroparticle Physics (TAsP-LNF).

\bibliographystyle{apsrev4-1}
\bibliography{main}{}

\end{document}